\documentclass[aps,12pt,pra]{revtex4}
\newcommand{\eqref}[1]{(\ref{#1})}
\def\sech{{\rm sech\,}}
\def\tanh{{\rm tanh\,}}
\def\cosh{{\rm cosh\,}}
\def\sinh{{\rm sinh\,}}
\begin{document}
\title{Quantum Hamilton-Jacobi analysis of PT symmetric
Hamiltonians} \author{S. Sree
Ranjani$^1$\footnote{akksprs@uohyd.ernet.in}, A.K.
Kapoor$^1$\footnote{akksp@uohyd.ernet.in}, and P. K.
Panigrahi$^{1,2}$\footnote{prasanta@prl.ernet.in}}
\address{$^1$ School of Physics, University of Hyderabad, Hyderabad-500
046, India\\ $^2$ Physical Research Laboratory, Navrangpura,
Ahmedabad-380 009, India}
\begin{abstract}
We apply the quantum Hamilton-Jacobi formalism, naturally defined
in the complex domain, to a number of complex Hamiltonians,
characterized by discrete parity and time reversal (PT) symmetries
and obtain their eigenvalues and eigenfunctions. Examples of both
quasi-exactly and exactly solvable potentials are analyzed and the
subtle differences, in the singularity structures of their quantum
momentum functions, are pointed out. The role of the PT symmetry
in the complex domain is also illustrated.
\end{abstract}
\maketitle

\section{Introduction}

Complex Hamiltonians possessing real eigenvalues have attracted
considerable attention in the current literature
\cite{bender,bender1,bender2,bender3,bender4,zno,kha,zaf,bag,lev,gl,oz,ba,can}.
These quantal systems, apart from being counter intuitive, are not
well understood because of their recent origin. These Hamiltonians
are characterized by discrete parity and time reversal symmetries.
In the case when the wave functions are also PT symmetric, the
eigenvalues are real and the violation of PT symmetry by the wave
function leads to eigenvalues, which occur in complex conjugate
pairs. Besides identifying new Hamiltonians belonging to this
class, the role of various discrete symmetries is also under
current investigation.

      The presence of complex potentials in these systems makes
them ideal candidates to be probed using the quantum
Hamilton-Jacobi (QHJ) formalism, since this approach has been
formulated in the complex domain \cite{lea,pad}. In the QHJ
formalism, the singularity structure of the quantum momentum
function (QMF) plays a crucial role in the determination of the
eigenvalues and the eigenfunctions. Recently, we have studied the
structure of the QMF in the complex domain for exactly \cite{sree}
and quasi-exactly solvable \cite{geo} potentials. In this light,
it is extremely interesting to investigate the properties of the
QMF of the PT symmetric Hamiltonians, in order to systematically
find out their differences and similarities with exactly solvable
(ES) real potentials, as also with the quasi-exactly solvable
(QES) ones \cite{Singh,Shifm,Ushve,atre}. It is worth mentioning
that, as compared to the solvable potentials, the QMF of QES
models reveal significant differences in their singularity
structure.

     The goal of this letter is to investigate the structure of the
QMF of a class of PT symmetric Hamiltonians, for which the
eigenfunctions and the eigenvalues are also simultaneously
obtained, through the QHJ approach. The differences and
similarities of these novel systems, with their ES and QES counter
parts, are clearly brought out.

\section{Quantum Hamilton-Jacobi formalism}

In this section, we proceed to describe briefly the QHJ formalism
and its working. More details can be found in our earlier paper
\cite{sree}. With the definition of QMF:

\begin{equation}
p =-i\hbar\frac{d}{dx}(\ln \psi),      \label{e1}
\end{equation}
the Schr\"{o}dinger eigenvalue equation $H\psi=E\psi$ can be cast
in the form of the Riccati equation,
\begin{equation}
p^2 -i\hbar p^{\prime} = 2m [E-V(x)].     \label{e2}
\end{equation}
To find the eigenvalues, Leacock and Padgett \cite{lea,pad}
suggested using the following exact quantization condition for the
bound states of a real potential:
\begin{equation} \label{quant}
\frac{1}{2 \pi} \oint_{C} pdx = n \hbar,     \label{e3}
\end{equation}
where, the contour $C$ encloses the $n$ moving poles in the
complex domain, corresponding to the nodes of the wave function
located in the classical region. This quantization rule is an
exact one and follows from the oscillation theorem in
Strum-Liouville theory. Using this quantization rule, Bhalla  {\it
et. al} \cite{bh1,bh4} studied several ES models and showed that
the eigenvalues could be obtained without obtaining the
eigenfunctions. Briefly, the integral in Eq.\eqref{e3} is
evaluated using the knowledge of the singular points of $p(x)$
outside the contour $C$ and their corresponding residues.

The singularities of $p$ consists of fixed and moving ones. The
fixed singular points of $p$ corresponding to the singular points
of the potential, can be found by inspection. From Eq.\eqref{e1},
one can see that the nodes of the wave function correspond to the
moving poles of the QMF. In general, the QMF may have other moving
poles at locations which are are not easy to determine. However,
the residues can be computed by substituting a Laurent expansion
in the Eq.\eqref{e2}. In fact, the residue at a moving pole can be
easily seen to be $-i\hbar.$

Application of the QHJ to PT symmetric potentials requires  an
approach different from the one used earlier. In the absence of a
generalization of the oscillation theorem \cite{bender2}, for this
class of potential, it is not clear whether the quantization rule
Eq.\eqref{quant}, is valid. Even in the case of the violation of
this quantization rule it is not clear which contour should be
used. Earlier studies indicate that, for all ES and QES models,
after a suitable change of variables, {\it the quantum momentum
function has a finite number of moving poles and the point at
infinity is at most a pole}. In this case the quantization
condition still holds for a contour, which encloses all the moving
poles. We shall assume this to be the case for the PT symmetric
models to be taken up in this paper.

When one attempts to compute the residue at a pole using the QHJ,
one gets two answers. A boundary condition, in the limit $\hbar
\rightarrow 0$, has been suggested by Leacock and Padgett to
select the right residue. Several other conditions, such as square
integrability, have also been utilized in earlier papers for this
purpose \cite{sree,geo}. In some cases \cite{akk,qpp}, in the
absence of any criterion to select a residue, one must consider
all the values which are consistent with the other equations of
the theory.

In studying the complex potentials, we will first perform a
suitable change of variable and bring the resultant equation to
the QHJ form so that the potential is replaced by a rational
function. We analyze the potential introduced by Khare and Mandal
in section 3 and the complex Scarf potential
 $ V(x) = -A \, \sech ^{2}x - iB \, \tanh x  \, \sech x $ with $A > 0$ in section 4.
 After comparing the structure of QMF for PT symmetric, ES and QES
 systems, we conclude in the final section, with the remarks about
 problems and future directions of work.

\section{Khare-Mandal model}

The potential expression, for the Khare-Mandal model is given by,
$V(x) = - (\zeta \cosh2x - iM)^2$. This potential has complex or
real eigenvalues depending on whether $M$ is odd or even
\cite{kha,bag}. It is worth noting that, parity operation in this
case is given by $x \rightarrow i\pi/2 -x$, whereas the time
reversal remains same as the conventional $i \rightarrow -i$.
Using the QHJ formalism, we first obtain the QES condition, for
the odd and even values of $M$. Subsequently, the explicit
expressions for the eigenvalues and eigenfunctions for the cases,
$M=3$ and $M=2$ are also obtained. The QHJ equation in terms of $q
\equiv ip$, setting $\hbar = 2m =1$, can be written as
\begin{equation}
q^2 + \frac{dq}{dx} + E+(\zeta \cosh2x - iM)^2 =0 .  \label{e4}
\end{equation}
This above form has the advantage that, the residue at each moving
pole is one. To bring the potential to the rational form we do a
change of variable $ t \equiv \cosh 2x$. Substitution of this in
Eq.\eqref{e4} gives
\begin{equation}
q^2 + 2 {\sqrt{t^2 - 1}}\, \frac{dq}{dt} + E + (\zeta t - i M)^2
=0. \label{e5}
\end{equation}
One observes that the coefficient of $\frac{dq}{dt}$ is not one.
Hence, in order to bring the above equation to the form of
Eq.\eqref{e4}, we define
\begin{equation}
q = 2(\sqrt{t^2 - 1})\phi   ,\quad \phi=\chi - \frac{t}{2(t^2 -1)}
\,\,  , \label{e6}
\end{equation}
which transforms \eqref{e5} to
\begin{equation}
\chi^2 +\frac{d\chi}{dt} +\frac{t^2 +2}{4(t^2 -1)^2} + \frac{E +
(\zeta
  t - iM)^2}{4(t^2 -1)}  =0  \,\, , \label{e7}
\end{equation}
which has the convenient form wherein the residue at a moving pole
is one. From here on, $\chi$ will be called as the QMF and
Eq.\eqref{e7}, the QHJ equation.

{\it Singularity structure of $\chi$: } As already explained in
the previous section, we shall assume that $\chi$ has a finite
number of moving poles in the complex $t$ plane and that the point
at infinity is at most a pole. Besides the moving poles, $\chi$
has fixed poles at $t = \pm 1$. It is seen from Eq.\eqref{e7} that
the function $\chi$ is bounded at $t=\infty$. Assuming that $\chi$
has only these above mentioned singularities, we separate the
singular part of $\chi$ and write it in the  following form:
\begin{equation}
\chi = \frac{b_1}{t-1}+ \frac{b^{\prime}_1}{t+1}+
\frac{P^{\prime}_n}{P_n} + C \, .   \label{e8}
\end{equation}
Here $b_1$ and $b^{\prime}_1$ are the residues at fixed poles
$t=\pm 1$; $P_n$ is a polynomial of degree $n$ and equals
$\prod_{k=1}^{n}(t-t_k)$, where $t_k$'s are the locations of the
moving poles of the QMF. $C$ gives the analytic part of $\chi$ and
is a constant due to the Liouville's theorem. From Eq.(7), one can
see that for large $t$, $\chi$ goes as $\pm \frac{i\zeta}{2}$,
which are the values of $C$.

   To find the residues at the fixed poles, say $t=1$, one needs to
expand $\chi$ in a Laurent series:

\begin{equation}
\chi = \frac{b_1}{t-1} +a_0 + a_1 (t-1) +\cdots  \, . \label{e9}
\end{equation}

Substituting this in Eq.\eqref{e7} and comparing coefficients of
different powers of $t$, one obtains the following two values for
$b_1$:
\begin{equation}
b_1 = \frac{3}{4} \, , \, \frac{1}{4}.      \label{e10}
\end{equation}
Similarly the two values of residues at $t=-1$ turn out to be,
\begin{equation}
b^{\prime}_1 = \frac{3}{4} \, , \, \frac{1}{4}.      \label{e11}
\end{equation}

{\it Behaviour at infinity :} It has been assumed that the point
at infinity is an isolated singularity. In order to find leading
behaviour of $\chi$ at infinity, one expands $\chi$ as,
\begin{equation}
\chi = a_0 +\frac{\lambda}{t} + \frac{\lambda_1}{t^2} +\cdots \,.
\label{e12}
\end{equation}
Substitution of Eq.\eqref{e12} which in Eq.\eqref{e7}, gives $
\lambda = \frac{iM \zeta }{4a_0}$ along with, $a_0 = \pm \frac{i
\zeta }{2}$, which is equal to $C$. Due to this $\lambda$ takes
the following two values:
\begin{equation}
\lambda = \frac{M}{2} \, , \, \frac{-M}{2}     \label{e14}
\end{equation}
This should match with the leading behaviour of $\chi$ coming from
Eq.\eqref{e8}, which is $\frac{b_1 +b^{\prime} +n}{t}$, for large
$t$. Hence equating the two equations, one obtains
\begin{equation}
b_1 + b^{\prime}_{1} + n = \lambda   \, .   \label{e15}
\end{equation}
From Eq.\eqref{e10} and \eqref{e11}, we see that the right hand
side of Eq.\eqref{e15} is positive. Hence for Eq.(14) to be true,
we choose only the positive value of $\lambda$ {\it i.e}, $\lambda
= \frac{M}{2}$, which means we choose $ a_0 = C =
+\frac{i\zeta}{2}$. It should be noted that, there is no way of
choosing a particular value of residue at a fixed pole, since one
does not have information regarding the square integrability of
the solutions. Hence, one needs to consider both the values of
$b_1$ and $b^{\prime}_1$. Thus taking all possible combinations of
$b_1$ and $b^{\prime}_{1}$ in Eq.\eqref{e14}, one obtains the QES
condition for each combination along with a constraint on $M$, as
given in table 1. From table 1, we see that sets 1 and 2 are valid
only when $M$ is odd and sets 3 and 4 are valid only when $M$ is
even.

{\it Forms of the wavefunction :} From Eq.\eqref{e1} one obtains
$\psi(x)$ is terms of $p$ with $\hbar = 2m=1$, as
\begin{equation}
\psi(x) = \exp (i\int pdx).  \label{e16}
\end{equation}
Doing the change of variable and writing $p$ in terms of $\chi$,
one gets,
\begin{equation}
\psi(t) = \exp \int \left( \frac{b_1}{t-1} +
\frac{b^{\prime}_1}{t+1} + \frac{P^{\prime}_n}{P_n} +\frac{i \zeta
}{2} - \frac{t}{2(t^2-1)} \right) dt.   \label{e17}
\end{equation}
Hence, one can substitute sets 1 and 2 in Eq.\eqref{e17} if $M$ is
odd and sets 3 and 4 if $M$ is even to obtain the form of the
wavefunction. The expression for the wave function is in terms of
the unknown  polynomial $P_n$, where $n$ gives the number of zeros
of $P_n$. In order to calculate the polynomial, we substitute
$\chi$ from Eq.\eqref{e8} in \eqref{e7}, to get
\begin{eqnarray}
\frac{P^{\prime\prime}_n}{P_n} + \frac{2P^{\prime}_n }{P_n}\left(
\frac{b_1}{t-1} + \frac{b^{\prime}_1}{t+1}+ \frac{i \zeta
}{2}\right) +\frac{b^{2}_1 - b_1}{(t-1)^2}
+\frac{(b^{\prime})^{2}_1 -
  b^{\prime}_1}{(t+1)^2} +\frac{t^2 +2}{4(t^2-1)^2} + \nonumber \\
\frac{E+(\zeta t
  -iM)^2 +8b_1 b^{\prime}_1 - 4\zeta  ^2 (t^2 -1)}{4(t^2 -1)} +i \zeta  \left(
\frac{b_1}{t-1} +\frac{b^{\prime}_1}{t+1}\right) =0.   \label{e18}
\end{eqnarray}
 This leads to $n$ linear homogeneous equations, for the coefficients of
 different powers of $t$ in $P_n$. The energy eigenvalues are obtained
 by setting the corresponding determinant equal to zero. The explicit
 eigenvalues and eigenfunctions are obtained for $M=3$ and $M=2$
 cases. \\
{\bf Case 1: M =3,}\\
    Here $M$ is odd so we can use sets 1 and 2 from table 1
 and get the required results as below.\\
{\it Set 1 :} $ b_1 =\frac{1}{4}$, $b^{\prime}_1 = \frac{1}{4}$
and
 $n=1$.\\
  This implies $P_n$ is a first degree polynomial say $Bt
 +D$. Substituting these values in Eq.\eqref{e18} and comparing various
 powers of $t$, one obtains a $2 \times 2$ matrix for $B$ and $D$ as
 follows

\begin{equation}
\left( \begin{array}{cc}
         1+ \frac{E-9+ \zeta  ^2}{4}    &  -i \zeta    \\
        -i \zeta        &  \frac{E-9+\zeta  ^2}{4}
\end{array}   \right)  \left( \begin{array}{c}
     B \\ D
\end{array}  \right)   = 0.    \label{e19}
\end{equation}
Equating the determinant of this matrix to zero, one obtains the
two values for energy and the polynomials as
\begin{equation}
E= 7- \zeta  ^2 \pm 2 \sqrt{1 - 4 \zeta  ^2}, \quad P_1 =
\frac{B}{2}(2t - \frac{i}{\zeta }(1 \pm \sqrt{1-4\zeta ^2})).
\label{e20}
\end{equation}
%which when substituted back in \eqref{e19}, gives back two values
%of $P_1$ as
%\begin{equation}
%P_1 = \frac{B}{2}(2t - \frac{i}{\zeta }(1 \pm \sqrt{1-4\zeta ^2}))
%\label{e21}
%\end{equation}
Substituting the values of $b_1$, $b^{\prime}_1$ and $P_1$ in
Eq.\eqref{e17} gives, the two eigenfunctions corresponding to the
two eigenvalues :
\begin{equation}
\psi(x) = e^{{\frac{i \zeta }{2}\cosh2x}}  \left(2\cosh2x
-\frac{i}{\zeta }(1 \pm
    \sqrt{1-4\zeta ^2})\right).    \label{e22}
\end{equation}
{\it set 2:} $ b_1 =\frac{3}{4}$, $b^{\prime}_1 = \frac{3}{4}$ and
 $n=0$.\\
Here we see that $n=0$ implies $P$ is a constant. Substituting
these
 values in Eq.\eqref{e18} and proceeding in the same manner as before one
 obtains,
\begin{equation}
E= 5- \zeta  ^2 ,\,\,\,  \psi(x) =e^{{\frac{i \zeta }{2} \cosh
2x}} \sinh2x \label{e23}
\end{equation}
which are the known results \cite{kha,bag}. Below we elaborate on
the case when $M$
is even.\\ {\bf Case 2: $M=2$,}\\
In this case, one makes use of sets 3 and 4 in table 1 and proceed
in
the same way as was done for case 1.\\
{\it set 3 :} $ b_1 =\frac{1}{4}$, $b^{\prime}_1 = \frac{3}{4}$
and
 $n=0$.\\
the eigenvalues and eigenfunctions obtained are
\begin{equation}
E = 3- \zeta ^2  +2i \zeta   ,\,\,\,   \psi(x) = e^{{\frac{i
      \zeta }{2}\cosh 2x}} (\cosh 2x +1)^{1/2}.    \label{e24}
\end{equation}

{\it set 4 :} $ b_1 =\frac{3}{4}$, $b^{\prime}_1 = \frac{1}{4}$
and
 $n=0$.\\
In this case, one obtains
\begin{equation}
E = 3- \zeta ^2  - 2i \zeta    ,\,\,\,   \psi(x) = e^{{\frac{i
      \zeta }{2}\cosh 2x}} (\cosh 2x -1)^{1/2}.    \label{e25}
\end{equation}
These match with the solutions given in \cite{bag}. Thus for any
given positive value of $M$, odd or even, one can obtain the
eigenvalues and eigenfunctions for the Khare-Mandal potential. In
the next section, we study the complex Scarf -II potential.

\section{Complex Scarf-II potential}

  The expression for the complex Scarf-II potential is given by
\begin{equation}
V = A \, \sech^{2}x + iB \, \sech x \, \tanh x.   \label{e26}
\end{equation}
Note that, unlike the previous case, here parity operation is
given by $x \rightarrow -x$, time reversal operation remaining the
same. The corresponding QHJ equation, in terms of $q$, where
$q=\frac{d \ln\psi}{dx}$, is
\begin{equation}
q^2 +\frac{dq}{dx} + E - A \sech^{2}x - iB \, \sech x \, \tanh x
=0. \label{e27}
\end{equation}
Carrying out the change of variable, $y= i \, \sinh x$, proceeding
in the same manner as before, one obtains the QHJ equation for
$\chi$ :
\begin{equation}
\chi^2 +\frac{d\chi}{dy} + \frac{2+y^2}{4(1-y^2)^2} -
\frac{E}{1-y^2} - \frac{A - B y}{(1-y^2)^2}  =0 \, ,   \label{e28}
\end{equation}
where
\begin{equation}
\chi = \left( \phi - \frac{y}{2(1-y^2)} \right) \, , \quad q = i
\, (\sqrt{1-y^2})\phi.   \label{e29}
\end{equation}
Along with the $n$ moving poles with residue one, $\chi$ has poles
at $y= \pm 1$. We assume that except for these poles there are no
other singularities and that the point at infinity is an isolated
singularity. The residue at $y=1$ and $-1$ are respectively given
by,
\begin{equation}
b_1 = \frac{1}{2} \pm \frac{1}{2}\sqrt{\frac{1}{4} + A -B},
\label{e30}
\end{equation}
and
\begin{equation}
b^{\prime}_1 = \frac{1}{2} \pm \frac{1}{2}\sqrt{\frac{1}{4} + A
+B} \, . \label{e31}
\end{equation}
As in Eq. (12), considering the behaviour of $\chi$ at infinity
one gets,
\begin{equation}
\lambda = \frac{1}{2} \pm \sqrt{-E}  \, .
\label{e32}\end{equation}

It should be noted that, unlike the previous QES case, energy
explicitly enters in $\lambda$. As will be soon seen, this is the
reason all the energy values can be obtained here, making this
exactly solvable model.

As seen in the earlier section one can write $\chi$ in terms of
its analytic and singular parts as
\begin{equation}
\chi = \frac{b_1}{y-1}+ \frac{b^{\prime}_1}{y+1}+
\frac{P^{\prime}_n}{P_n} + C  \, ,  \label{e33}
\end{equation}
where $C$ is the analytic part of $\chi$. $C$ is a constant due to
Liouville's theorem, which turns out to be zero. The leading
behaviour of $\chi$ for large $y$ from Eq.\eqref{e33} is of the
form, $\frac{b_1 +b^{\prime}_1 +n}{y}$. This coefficient of
$\frac{1}{y}$, should match with $\lambda$ in Eq.(30) {\it i.e.},
\begin{equation}
b_1 +b^{\prime}_1 +n = \lambda.     \label{e34}
\end{equation}
This gives the energy eigenvalues as
\begin{equation}
-E = (b_1 +b^{\prime}_1 +n -\frac{1}{2})^2\, .   \label{e35}
\end{equation}
The wave function  in terms of $\chi$ is
\begin{equation}
\psi(y) = \exp \left( \int dy \left(
\frac{b_1}{y-1}+\frac{b^{\prime}_1}{y+1} +\frac{P^{\prime}_n}{P_n}
+ \frac{1}{2}\frac{y}{1-y^2} \right)\right) ,
  \label{e36}
\end{equation}
which is equal to
\begin{equation}
\psi(y) = (y-1)^{-p} (y+1)^{-q} P_n(y),    \label{e37}
\end{equation}
where, $-p = b_1 -\frac{1}{4}$ and $ -q = b^{\prime}_1
-\frac{1}{4}$. To obtain the polynomial, one needs to substitute
Eq.\eqref{e33} in \eqref{e28}, which yields the following
differential equation
\begin{equation}
P^{\prime\prime}_n + 2P^{\prime}_n \left(\frac{b_1}{y-1} +
\frac{b^{\prime}_1}{y+1}\right) + G(y)P_n = 0,    \label{e38}
\end{equation}
where
\begin{eqnarray}
G(y) = \frac{(4(b^2 _1 -b_1 +b^{\prime 2}_1-b^{\prime}_1)) +1
+4E)y^2 + 2y(4(b^2 _1 -b_1 -b^{\prime 2}_1+b^{\prime}_1)
+2B))}{(y^2-1)^2}+ \nonumber \\
\frac{(4(b^2 _1 -b_1 +b^{\prime 2}_1 -b^{\prime}_1) +2 - 4A -4E)}{(y^2-1)^2}. \nonumber\\
\end{eqnarray}
 Substituting the expression for $E$ from Eq.\eqref{e35}
 in \eqref{e38}, one obtains
\begin{equation}
(1-y^2)P^{\prime\prime} + P^{\prime} (2(b_1 - b^{\prime}_1) - 2
(b_1 +
  b^{\prime}_1)y) +n(n+2(b_1 +b^{\prime}_1 -1) +1 ) =0 \, ,   \label{e39}
\end{equation}
which is in the form of the Jacobi differential equation and hence
the polynomial $P_n (y) = P^{2b_1 -1, 2b^{\prime}_1 -1} _ n (y)$
is the Jacobi polynomial. Thus the complete expression for the
wave function can be written as,
\begin{equation}
\psi(x) = (i \, \sinh x -1)^{b_1 - \frac{1}{4}} (i \, \sinh x
+1)^{b^{\prime}_1
    -\frac{1}{4}} P ^{\,\, 2b_1 -1, 2b^{\prime}_1 -1} _n (i \, \sinh x),    \label{e40}
\end{equation}
which matches with the answer in \cite{zaf} if $b_1$ and
$b^{\prime}_1$ are written in terms of $p$ and $q$ respectively.
Note that in this whole process, we had written the expression of
the eigenvalues and the eigenfunctions in terms of $b_1$ and
$b^{\prime}_1$ which have two values. No particular value has been
chosen. Hence, we need to choose one value of each residue to
remove this ambiguity. For this case, the  solutions are known to
satisfy the property $\psi(\pm \infty) \rightarrow 0 $. We make
use of this condition to choose the right values of the residues.
For this purpose, we consider two conditions on the potential
parameter $A$ and $B$. For each condition, we see that the
particular values of residues, which are chosen using the property
of square integrability give physically acceptable results.

{\bf Case 1: $ |B| > A + \frac{1}{4}$}

With this restriction on $A$ and $B$, the residues at $y = \pm 1$
becomes, $b_1 = \frac{1}{2} \pm \frac{i}{2}\sqrt{B-A-\frac{1}{4}}$
and $b^{\prime}_1 = \frac{1}{2} \pm
\frac{1}{2}\sqrt{A+B+\frac{1}{4}}$. In the limit $y \rightarrow
\infty$, the wave function in Eq.(35) goes as
\begin{equation}
\psi(y) \approx y^{b_1 +b^{\prime}_1 -\frac{1}{2} +n}. \label{e42}
\end{equation}
The above equation, with the values of $b_1$ and $b^{\prime}_1$
substituted becomes
\begin{equation}
\psi(y) \approx y^{\pm
\frac{i}{2}\sqrt{B-A-\frac{1}{4}}}y^{\frac{1}{2} \pm
  \frac{1}{2} \sqrt{A+B+\frac{1}{4}} +n}     \label{e43}
\end{equation}
For $\psi(y)$ to go to zero for large $y$, $b^{\prime}_1 =
\frac{1}{2} + \frac{1}{2}\sqrt{A+B+\frac{1}{4}}$ is ruled out.
Hence the choice of the residues for this case will be
\begin{equation}
b_1 = \frac{1}{2} \pm \frac{i}{2}\sqrt{B-A -\frac{1}{4}} \,\,\,\,
\,\,\,\,\, b^{\prime}_1 = \frac{1}{2} -
\frac{1}{2}\sqrt{A+B+\frac{1}{4}},   \label{e44}
\end{equation}
with the restriction on $n$ as
\begin{equation}
n < \frac{1}{2}\sqrt{A+B+\frac{1}{4}} - \frac{1}{2}. \label{e44a}
\end{equation}
For these values of the residues at the fixed poles, $\psi$ takes
the form,
\begin{equation}
\psi = (i \sinh x -1)^{\frac{1} {4} \pm \frac{ir}{2}} (i \sinh x
  +1)^{\frac{1}{4} -\frac{s}{2}} P^{\pm ir, s}_n (i \sinh x),    \label{e44}
\end{equation}
where $r = \sqrt{B-A-\frac{1}{4}}$ and $s =
\sqrt{A+B+\frac{1}{4}}$. The expression for energy is
\begin{equation}
E = -\left( n+\frac{1}{2} -\frac{1}{2}
\left(\sqrt{A+B+\frac{1}{4}} \pm i \sqrt{B-A-\frac{1}{4}} \right)
\right ) ^2,     \label{e45}
\end{equation}
with the condition on $n$ as
\begin{equation}
n < \frac{1}{2}\sqrt{\frac{1}{4} +A -B } +
\frac{1}{2}\sqrt{A+B+\frac{1}{4}} - \frac{1}{2}.    \label{e46a}
\end{equation}

{\bf Case 2 : $|B| \le A +\frac{1}{4}$}

Proceeding in the same way as above, one obtains the choice of the
residues as
\begin{equation}
b_1 = \frac{1}{2}- \frac{1}{2}\sqrt{\frac{1}{4} +A-B}  ,\,\,
b^{\prime}_1 = \frac{1}{2}- \frac{1}{2}\sqrt{\frac{1}{4} +A+B}.
\label{e46}
\end{equation}
The wave function is given by,
\begin{equation}
\psi = (i \sinh x -1)^{\frac{1}{2} -\frac{\mu}{2}} (i \sinh x
+1)^{\frac{1}{2} - \frac{\nu}{2}} P^{-\mu,\nu}_n (i \sinh x),
\label{e47}
\end{equation}
where $\mu = \sqrt{\frac{1}{4} +A -B}$ and $ \nu =
\sqrt{\frac{1}{4} +A
  +B}$ with the energy
\begin{equation}
E = -\left( n+\frac{1}{2} -\frac{1}{2}\left
(\sqrt{\frac{1}{4}+A-B} + \sqrt{\frac{1}{4}+A+B} \right ) \right )
^2  .   \label{e45}
\end{equation}
Thus the answers match with those given in \cite{zaf}.

\section{Conclusions}

In conclusion, PT symmetric potentials belonging to the QES and ES
class have been investigated through the QHJ formalism. The QES
solvable Khare-Mandal potential has complex or real eigenvalues,
depending on whether the potential parameter $M$ is odd or even.
The singularity structure of the QMF for these two cases is
different.  For the case when $M$ is odd, one observes from table
I that the solutions fall into two groups, which consist of
solutions coming from sets 1 and 2. For a solution belonging to a
particular group, the number of singularities of the QMF are fixed
and consist of both real and complex locations. This kind of
singularity structure of the QMF has been observed in the study of
periodic potentials \cite{akk}. Though the solutions, for $M$
even, fall into two groups coming from sets 3 and 4,  they  all
have same number of singularities, which again can consist of
complex and real poles. This singularity structure is same as
observed in the ordinary QES models \cite{geo}.

Coming to the case of exactly solvable PT symmetric potential, the
location of the moving poles can be either real or complex. In the
specific example of complex Scarf potential, it turns out that all
the moving poles are off the real line. In contrast for the
ordinary ES models the moving poles are always real. For both the
cases, the number of moving poles of the QMF, characterize the
energy eigenvalues.

%-----------------------------------------------------------

{\bf{Acknowledgements}}: We thank Dr. Z. Ahmed for illuminating
 discussions and Rajneesh Atre for carefully going through the
 manuscript.

 \begin{table}%[p]
\caption{This table gives the QES condition and the number of
moving
  poles of $\chi$ for each combination of $b_{1}$ and $b^{\prime}_1$
  for the Khare-Mandal model  }
\vskip0.5cm
\begin{tabular}{ c c c c c c  }
\hline
%\multicolumn{7}{|c|}
%   \\  \hline
\multicolumn{1}{c}{set} &\multicolumn{1}{c}{$b_1$}
&\multicolumn{1}{c}{$d_1$} &\multicolumn{1}{c}{$n = \lambda -b_1 -
b^{\prime}_1$} &\multicolumn{1}{c}{Condition on M}
&\multicolumn{1}{c}{QES condition}
\\ \hline
  &              &              &                             &  & \\
1 & $\frac{1}{4}$ & $\frac{1}{4}$ & $\frac{M}{2}-\frac{1}{2}$ & $
M =
odd,\, M \ge 1 $    & $M =2n+1$  \\
2 & $\frac{3}{4}$ & $\frac{3}{4}$ & $\frac{M}{2}-\frac{3}{2}$ & $
M =
odd,\, M  \ge 3 $   & $ M= 2n +3$ \\
3 & $\frac{3}{4}$ & $\frac{1}{4}$ & $\frac{M}{2}-1$ &$ M =
even,\, M \ge 2$    & $M = 2n +1$ \\
4 & $\frac{1}{4}$ & $\frac{3}{4}$ & $\frac{M}{2}-1$           &$ M
=
even,\, M \ge 2$    & $M = 2n +1$ \\
%  &               &               &                           &  &  \\
\hline
\end{tabular}
\end{table}

\end{document}